\documentclass[11pt]{article} 
\usepackage{amsfonts,amscd,amsmath}
\newcommand{\oh}{\frac{1}{2}}

\def\ep{\text{e}}
\def\as{\alpha_s}
\def\ap{\alpha_{\text r}}
\def\X{\mathbf{X}}
\def\g{\mathfrak{g}}

\title{$1/q^2$  Corrections and Gauge/String Duality}
\author{Oleg Andreev\thanks{Also at Landau Institute for Theoretical Physics, Moscow, Russia. Email 
address: andre@itp.ac.ru}
\\ \\
{\it Max-Planck Institut f\" ur Physik, F\" ohringer Ring 6,} \\
{\it 80805 M\" unchen, Germany}}
\date{}

\begin{document} 

\vspace{-8cm} 
\maketitle 
\begin{abstract} 
We make an estimate of the quadratic correction based on gauge/string duality. Like in QCD, it proves to be negative and 
proportional to the string tension.
 \end{abstract}

\vspace{-10cm}
\begin{flushright}
MPP-2006-31
\end{flushright}
\vspace{9cm}


\section{Introduction}

As is well known, the QCD analysis of the two current correlator results in 

\begin{equation}\label{2currents}
i\int d^4x\,\ep^{iq\cdot x}\langle \,\text{T}J^\mu(x) J^\nu(0)\,\rangle =\left(q^\mu q^\nu -\eta^{\mu\nu}q^2\right)\Pi (q^2)
\,,
\end{equation}
with
\begin{equation}\label{ptensor}
{\cal N} q^2\,\frac{d \Pi }{dq^2}=C_0+\frac{1}{q^2}C_2+
\sum_{n\geq 2}\frac{n}{q^{2n}}C_{2n}\langle {\cal O}_{2n}\rangle
\,.
\end{equation}

\noindent Here $\eta_{\mu\nu}$ is a four-dimensional Minkowski metric, $q\cdot x\equiv q_\mu x^\mu$, and ${\cal N}$ is some 
normalization factor. We take $\eta_{\mu\nu}=\text{diag}(-1,1,1,1)$ to connect with standard string calculations. 

According to \cite{svz}, $C_0$ is a coefficient that can be calculated perturbatively in $\as$. So, it is of the form 
$C_0=1+\sum_{n\geq 1}B_n\as^n(q^2)$. The non-perturbative effects are generated by {\it local} gauge invariant operators 
whose dimensions $D$ are larger or equal four. For example, the $D=4$ operators  built from the quark and 
gluon fields are simply ${\cal O}_4^M=\bar qMq$ and ${\cal O}_4^G=\as G^2$. It is clear that in such an approach the 
existence of the quadratic correction $\tfrac{1}{q^2}C_2$ is puzzling. Later \cite{q2}, this issue got intensively discussed in the 
literature. In particular, there are estimates of the quadratic correction based on the data for the $e^+e^-$ total cross section that 
provide the upper bound \cite{data}

\begin{equation}\label{C2}
|\,C_2\,|\leq 0.14 \,\text{GeV}^{\,2}
\,.
\end{equation}

One of the implications of the AdS/CFT correspondence is that it resumed interest in finding a string description of strong 
interactions. For the case of interest, let us briefly mention two approaches.  In the first, usually called AdS/QCD,  one starts from a
five-dimensional effective field theory somehow motivated by string theory and tries to fit it to QCD as much as possible. Such an 
approach yields the leading asymptotic piece $C_0$ \cite{son} and even condensates of the operators with $D\geq 4$ \cite{hirn}. In 
the second, usually called gauge/string duality, one tries to keep the underlying string structure. As a result, the theory is ten 
dimensional and its reductions to five dimensions in general contain additional higher derivative 
terms ($\alpha'$ corrections).\footnote{These corrections are of order $\tfrac{1}{\sqrt{N_c}}$. Thus, they might be relevant 
at $N_c=3$.} In the approximation of \cite{joe}, this approach also yields the leading asymptotic piece $C_0$ \cite{oa}.

In this note we address the issue of the quadratic correction within the simplified model of \cite{joe,oa}. We compute them by 
following the strategy of first quantized string theory. To this end, we build the corresponding vertex operators. Then, we define 
the correlator of two vector currents as an expectation value of the vertex operators. We also adopt the geometric 
approach to condensates introduced in \cite{hirn}. Our aim is to estimate the correction and compare the result with that of QCD.  

Before proceeding to the detailed analysis, let us set the basic framework. We consider the following background metric

\begin{equation}\label{metric}
ds^2=\tfrac{R^2}{r^2}h
\Bigl(\eta_{\mu\nu}dx^\mu dx^\nu+dr^2\Bigr)+g_{ab}\,d\omega^ad\omega^b
\,,\quad\quad h=\ep^{-\oh cr^2}\,,
\end{equation}
where $\omega^a$ are coordinates of some five dimensional compact space $\text{X}$. In the region of small $r$ the metric 
behaves asymptotically as ${\text AdS}_5\times\text{X}$, as expected. We take a constant dilaton and, unfortunately, discard all 
possible RR backgrounds (if any). 

As known, full control of superstring theory on curved backgrounds even like $\text{AdS}_5$ is beyond our grasp at present. We 
are forced therefore to look for a plausible approximation. The simplest possible one is that of \cite{joe}. The idea is to discard nonzero 
modes of sigma model fields $\mathbf{r}$ and $\boldsymbol{\omega}^a$.\footnote{$\mathbf{X}$, $\mathbf{r}$ and 
$\boldsymbol{\omega}$ are taken to be sigma model fields on a string worldsheet. $x$, $r$ and $\omega$ are their zero modes, 
respectively.} As a result, the $\mathbf{X}$'s only contribute to the kinetic terms of the worldsheet action which is of the form 

\begin{equation}\label{action}
S_0=\frac{1}{4\pi\ap}\int_{\Sigma}d^2zd^2\theta\,\eta_{\mu\nu}\bar D\X^\mu D\X^\nu \,,\quad
\ap=\alpha' \tfrac{r^2}{R^2}h^{-1}\,.
\end {equation}
Here $\X$ is a two-dimensional superfield and $\Sigma$ is a two-dimensional Riemann surface.\footnote{We use the superspace 
notations of \cite{fms}.} 

Finally, a couple of remarks are in order:

First, the action \eqref{action} formally looks like that in flat space. So, it is conformally invariant. The point is, however, 
that the string parameter is now a variable and, as a consequence, the theory behaves differently.

Second, even such a simplified approximation contains $\alpha'$ corrections and remnants of the compact space $\text {X}$.

\section{Estimate of the Quadratic Correction}

As a warmup, let us fix the parameter $c$. A possible way to do so is to consider meson operators. As usual in first quantized string 
theory, we should look for the corresponding vertex operators. According to \cite{oa}, these are of the form 
\begin {equation}\label{meson}
{\cal O}(\xi,p)=\oint_{\partial\Sigma} dz d\theta \,
\xi\cdot D\mathbf{X} \,\Psi_p(\mathbf{X},\mathbf{r},\boldsymbol{\omega})
\,,
\end{equation}
where $\xi_\mu$ is a polarization vector. $\partial\Sigma$ denotes a worldsheet boundary. $\Psi_p$ is a solution 
to the linearized ten-dimensional Yang-Mills equations in the 
background \eqref{metric}. We choose the simplest possible solution that is $\Psi_p=\ep^{ip\cdot\mathbf{X}}\psi(\mathbf{r})$ 
with $\psi$ being a solution to the following equation\footnote{ Since we discard the nonzero modes of $\mathbf{r}$ and 
$\boldsymbol{\omega}$, we set the corresponding YM connections to be zero. Moreover, in such an approximation there is 
gauge invariance $A_\mu\rightarrow A_\mu+\partial_\mu\Lambda$. It is fixed by $\partial\cdot A=0$.} 

\begin{equation}\label{Lag}
r^2\psi''-r\left(1+\tfrac{1}{2}cr^2\right)\psi'+m^2r^2\psi=0
\,,
\end{equation}
where $m^2=-p\cdot p$. A prime denotes a derivative with respect to $r$.

It is easy to reduce Eq.\eqref{Lag} to Laguerre differential equation whose solutions are given by Laguerre polynomials 
$L_n^a$. For 

\begin{equation}\label{mass}
m^2=cn
\,,
\end{equation}
 we get

\begin{equation}\label{Lag2}
\psi_n(r)=c r^2 L^1_{n-1}\left(\tfrac{1}{4}cr^2\right)
\,,\quad\text{with} 
\quad n=1,\,2,\,\dots
\,.
\end{equation}

Since the mass of the $n$-th state is proportional to $\sqrt{n}$, it seems natural to interpret the vertex operators \eqref{meson} as 
those of $\rho (m_n)$ mesons. If so, then the value of $c$ is of order \cite{pdg}

\begin{equation}\label{sigma}
c\approx 0.9 \,\text{GeV}^{\,2}\,.
\end{equation}

We close the discussion of the meson operators with a few short comments:
\newline (i)  The comparison of the mass formula \eqref{mass} with the real meson masses shows a better fit for 
large $n$.\footnote{In fact, starting from $n=3$ the agreement is better than $10\%$.} Evidently the parameter $c$ is proportional 
to the string tension. 
\newline (ii) By contrast, the metric used in \cite{hirn} to reproduce the operators with $D\geq 4$ doesn't lead to a simple equation 
yielding the exact mass formula. 
\newline (iii) It is worth mentioning that the mass formula \eqref{mass} has been derived in the $\text{AdS}_5$ background with 
a non-constant dilaton $\Phi $ \cite{katz}

\begin{equation}\label{katz}
h=1\,,\quad\quad
\Phi (r)=\tfrac{1}{2}cr^2
\,.
\end{equation}
This background is indeed equivalent to ours as long as one deals with {\it quadratic} terms like $F^2$ in $5$-dimensional 
effective actions because the Weyl rescaling of the metric transforms one into another. In the generic case, the equivalence is broken by 
higher order terms in the field strength or by scalar fields. 

Having fixed the value of the parameter $c$, let us make an estimate of the quadratic correction. To this end, we compute the 
two current correlator.

We start with a vertex operator construction for the vector currents. In the spirit of \cite{oa}, we can take the operators to be 

\begin {equation}\label{current}
J^\mu (q)=\oint_{\partial\Sigma} dz d\theta \,
D\mathbf{X}^\mu \,\Psi_q(\mathbf{X},\mathbf{r},\boldsymbol{\omega})
\,,
\end{equation}
with the simplest possible form for $\Psi_q$ which is $\ep^{iq\cdot\mathbf{X}}\psi(\mathbf{r})$. Again, $\psi$ is a solution to a 
differential equation. This equation is given by \eqref{Lag} with $m^2=-q^2$. 

Next, we define the correlator as an expectation value of the vertex operators. In doing so, we choose a unit disk as the worldsheet.
Using the result of \cite{oa}, it is easy to write down the following expression\footnote{We normalize the vertex 
operators \eqref{current} in such a way to fit \eqref{ptensor}.}

\begin{equation}\label{ptensor2}
{\cal N}q^2\,\frac{d \Pi }{dq^2}=-2\int_0^1 dz\int_0^{\infty}dt\,h^{\frac{1}{2}}
\Bigl
(\psi\psi'+\g t\psi^2 h^{-1}G(z)
\Bigr)
\exp\Bigl\lbrace  \g t^2 h^{-1}G(z)\Bigr\rbrace
\,,
\end{equation}
where $t=qr$, $q=\sqrt{q\cdot q}$, $h=\ep^{-\tfrac{1}{2}\lambda t^2}$, $\lambda=\tfrac{c}{q^2}$, and $\g =\tfrac{\alpha'}{R^2}$. 
$G(z)$ denotes a restriction of the scalar Green function on the boundary.\footnote{ As follows from \eqref{action}, 
it is given by $\langle X^\mu(z)X^\nu(0)\rangle =\ap\eta^{\mu\nu}G(z)$.} Note that it excludes the zero mode contribution.

Before continuing our discussion, we pause here to take a closer look at \eqref{ptensor2}. A natural question to ask is to what extent 
this expression can be used as a basis for providing the $C_{2n}$'s. First, having restricted ourselves to the disk topology, we perhaps 
discard perturbative $\alpha_s$ corrections.\footnote{The line of thought that perturbation theory at hand is a topological expansion was 
pursued in \cite{oa2, oa}.} Second, in the approximation under consideration the integrand of \eqref{current} is not a dimension one half 
operator for generic $q$. As follows from \eqref{action}, its dimension is given by $\Delta=\tfrac{1}{2}+O(\ap q^2)$. It is clear that 
the problem is due to exclusion of the nonzero modes of $\mathbf{r}$.  Thus, it seems reasonable to restrict ourselves in 
\eqref{ptensor2} to a few leading terms in $\g$. Finally, $G(z)$ is singular at $z=0$. To deal with this divergence, we regulate the Green 
function as \cite{ats}

\begin{equation}\label{regul}
G(z)=2\sum_{n=1}^\infty \frac{\ep^{-\varepsilon n}}{n}\cos 2\pi nz
\,,
\end{equation} 
where $\varepsilon$ is some parameter.

Expanding in $\g$ and keeping only the constant and linear terms, we find

\begin{equation}\label{ptensor3}
{\cal N}q^2\,\frac{d \Pi }{dq^2}=-2\int_0^{\infty}dt
\,h^{\frac{1}{2}}\psi\psi'
\,.
\end{equation}
 Note that the linear term vanishes as a consequence of $\int_0^1 dz \,G(z)=0$.

Now the coefficients $C_{2n}$ can be read off from \eqref{ptensor3}. For the case of interest, we have

\begin{equation}\label{C02}
C_0=1\,,\quad\quad
C_2=-\tfrac{1}{3}c
\,.
\end{equation}
For completeness, we have included a calculation of \eqref{C02} in the appendix.

Finally, we substitute \eqref{sigma} into \eqref{C02} and obtain the following estimate

\begin{equation}\label{strC2}
C_2\approx - 0.3\,\text{GeV}^{\,2}
\,.
\end{equation}
This is our main result. The sign of $C_2$ is precisely as in QCD but the absolute value is at least twice bigger than the upper 
bound \eqref{C2}. 

\section{Concluding Comments}
(i) Having derived the mass formula \eqref{mass}, it seems to be time to estimate the Regge parameters. As noted above, this formula 
is more or less appropriate for $n\geq 3$. Assuming that all the trajectories of interest are parallel \cite{man}, we get for the intercepts
\begin{equation}\label{int}
\boldsymbol{\alpha}_n(0)\approx-\tfrac{c}{2\pi\sigma} n
\,,\quad\quad n=3,4,\dots
\,,
\end{equation}
where $\sigma$ is the string tension. 

As noted earlier, the mass formula is in considerable disagreement with the $\rho$ trajectory ($n=1$). This means that the background 
we used is not exactly the desired string dual to QCD and we should look for a further refinement of it. From this point of view the 
disagreement between our estimate and QCD is not a big surprise.
\newline (ii) Here we used the model with the slightly deformed $\text{AdS}_5$ metric. It is clear that this is 
not the only option. For example, the other line of thought is to deform the dilaton \cite{katz}.\footnote{Perhaps, both the models look 
oversimplified but they might be useful in gaining some intuition about the problem, while full control of superstring theory on 
curved backgrounds is missing.} What really fits better to QCD remains to be seen.
\newline (iii) Interestingly enough, it was argued by Zakharov that $C_2$ represents the stringy effect \cite{vi}. This fact being far 
from obvious in QCD is manifest in our framework. 

\vspace{.25cm}
{\bf Acknowledgments}

\vspace{.25cm}
We would like to thank V.I. Zakharov and P. Weisz for useful and stimulating discussions, and also to thank 
S. Brodsky and G. de Teramond for comments on the manuscript. This  work  is supported in part by the Alexander von Humboldt 
Foundation and Russian Basic Research Foundation Grant 05-02-16486. We are grateful for the hospitality extended to us at 
the Heisenberg Institut in M\" unchen.

\vspace{.35cm} 
{\bf Appendix}
\renewcommand{\theequation}{A.\arabic{equation}}
\setcounter{equation}{0}

\vspace{.25cm} 
\noindent In this appendix we present a calculation of \eqref{C02}.

Having derived the integral representation \eqref{ptensor3}, we can analyze its expansion in $\lambda$ that is nothing else but 
the expansion \eqref{ptensor}. For the coefficients of interest, we have

\begin{equation}\label{C02-1}
C_0=-\psi_0^2\,\Bigl\lvert_{t=0}^\infty
\,,\quad\quad
C_2=\tfrac{1}{4}c\left(t^2\psi_0^2-8\psi_0\psi_1\right)\Bigl\lvert_{t=0}^\infty-
\tfrac{1}{2}c\int_0^{\infty}dt\,t\psi_0^2
\,,
\end{equation}
Here we have expanded $\psi$ as a series in powers $\lambda$ such that $\psi=\sum_{n=0}\psi_n\lambda^n$. Notice that the 
$\psi_n$'s obey a set of differential equations. In particular, $\psi_0$ and $\psi_1$ are determined from 

\begin{equation}\label{diff}
t\psi_0^{''}-\psi_0^\prime-t\psi_0=0
\,,\quad\quad
t\psi_1^{''}-\psi_1^\prime-t\psi_1=\tfrac{1}{2}t^2\psi_0^\prime
\,.
\end{equation}
We also impose the following boundary conditions $\psi(0)=1$ and $\psi(\infty)=0$ that in one turn provide

\begin{equation}\label{boun}
\psi_0(0)=1\,,
\quad\quad 
\psi_n(0)=0\,,
\quad\quad\text{for}\quad\quad n\geq 1
\,,\quad\quad
\psi_n(\infty)=0
\,,\quad\quad 
\forall n
\,.
\end{equation}
 Given the boundary conditions,  the appropriate solutions to \eqref{diff} are simply given by 

\begin{equation}\label{psi0}
\psi_0 (t)=t K_1(t)
\,,\quad\quad
\psi_1(t)=\tfrac{1}{8}t^3 K_1(t)
\,.
\end{equation}
The remaining integral may be found in tables \cite{gr}. It is given by

\begin{equation}\label{intK}
\int_0^{\infty}dt\,t^3 K_1^2(t)=\tfrac{2}{3}
\,.
\end{equation}

So finally, we get

\begin{equation}\label{C02-2}
C_0=1\,,
\quad\quad
C_2=-\tfrac{1}{3}c
\,.
\end{equation}


\end{document}